\documentclass[aps,epsf,preprint]{revtex4}
\usepackage{amssymb}
\usepackage{amsmath}
\usepackage{color}

\def\[{\left[}
\def\]{\right]}
\def\be{\begin{eqnarray}}
\def\ee{\end{eqnarray}}
\def\bm{\begin{pmatrix}}
\def\em{\end{pmatrix}}
\def\ba{\begin{array}}
\def\ea{\end{array}}
\def\nn{\nonumber}
\def\({\left(}
\def\){\right)}
\def\bk#1{\langle#1\rangle}
\def\eq#1{Eq.(\ref{#1})}
\def\a{\alpha}
\def\s{\sigma}
\def\e{\epsilon}

\def\f{\phi}

\def\l{\lambda}
\def\m{\mu}
\def\x{\times}

\def\d{\delta}
\def\labels#1{\label{#1}}
\def\bn{\begin{enumerate}}

\def\en{\end{enumerate}}
\def\b{\beta}
\def\g{\gamma}
\def\ba{\begin{array}}
\def\ea{\end{array}}
\def\bc{\begin{center}}
\def\ec{\end{center}}

\def\.{\!\cdot\!}
\def\igw#1{\includegraphics[width=#1cm]}
\def\+{\!+\!}
\def\-{\!-\!}
\def\vs{\vskip.5cm}

\def\M{M\"obius\ }

\def\pf{{\rm Pf}}
\def\r{\rho}
\def\h{{1\over 2}}

\def\Mb{{\overline M}^2}
\def\mp{{m'}}
\def\mpp{{m''}}
\def\n{\nu}
\def\sh{\sqcup\mathchoice{\mkern-7mu}{\mkern-7mu}{\mkern-3.1mu}{\mkern-3.8mu}\sqcup}

\usepackage{amssymb}
\usepackage{graphicx}

\begin{document}
\title{CHY Theory for Several Fields}
\author{C.S. Lam}
\affiliation{Department of Physics, McGill University\\
 Montreal, Q.C., Canada H3A 2T8\\
Department of Physics and Astronomy, University of British Columbia,  Vancouver, BC, Canada V6T 1Z1 \\
Email: Lam@physics.mcgill.ca}
\title{CHY Theory for Several Fields}

\begin{abstract}
The Cachazo-He-Yuan (CHY) formula was originally proposed to describe  on-shell scattering of 
particles from a single massless field. We present a method to modify it to include
several interacting scalar fields, all possessing different masses and possibly off-shell momenta. 
The method is applied to 
Yukawa interactions between a number of scalar nucleons and pions,
 and to the $\f_1\f_2\f_3$ coupling of three different scalar fields.  Composite models  constructed from 
existing theories can be used to broaden the scope of the method.  The modification is  applied to describe Compton scattering from a massive particle, and to photon bremsstrahlung. It is also employed to generalize the disk function $Z$ and the sphere function $J$. 
\end{abstract}

\maketitle

\section{Introduction}
Inspired by string theory in the  zero Regge slope limit, 
Cachazo-He-Yuan (CHY) came up with a formula for the tree amplitude of several massless field 
theories, valid  in any number of space-time dimension \cite{CHY13a, CHY13b, CHY13c}. Like the string theory, the
CHY formula is given by  a multiple integral over the complex plane, possessing \M invariance. A set of {\it scattering equations}
plays an important role in this formalism. Propagators are now closely related to the 
scattering functions in the scattering equations,
and vertices are tied up with the global structure of the amplitude.

As  originally conceived,
the CHY formula describes the scattering of a massless scalar field with $\f^3$ coupling, the pure Yang-Mills theory, and Einstein's gravity. These are all massless
fields with massless external momenta. Because the formula is applicable in any number of space-time dimension,
dimensional reduction can be used to construct many other theories in 
lower dimensions \cite{CHY14a, CHY14b}. 

The correctness of the CHY formula was proved in \cite{DG13} and  directly verified by computations \cite{CG15, BBBD15a, BBBD15b, BBBDF15, LY15b,LY16, HFLZ16, BBDF16, HZ16b, CCWX17, HDF17, GHZ17, Lam18}. Properties and solutions
of the scattering equations have been investigated \cite{ Kal13, Wein14a, DG14, HMS14, Nau14a, Lam14, HRFH15, SZ15, CK15a, CK15b, DG15, DTW16, Zlo16, CMZ16, Miz17}, soft and collinear limits have been derived \cite{SV14,CHY15a,CCM16,Sah16, DPW16}, and loop computations have been
attempted \cite{HY5, BBBCDF15, GMMT15, CHY15b, Fen16, GMMT16, Gom17, GLT17}. It has been reformulated as a string theory \cite{MS13, GLM14}, and as a double-cover integral \cite{Gom16}. The relations between different CHY theories have been studied using differential operators  and scattering forms \cite{CSW17, CKW17, ABHY17, HYZZ18, HZ18, ZF18, BF18, FLZ19, NR19}. The connection between CHY and string theories
has also been investigated \cite{BSS13, MZ17, HTZ18, HTZ19, HRZ20}.

As interesting and as novel as this new formulation is, it would not be  helpful to high energy physics unless it can 
describe the Standard Model. That would require a number of additional developments not contained in the 
original formula.
The formulation must be broadened to include off-shell amplitudes in order  to coincide with field theory, and for loops to be computed.
Non-zero masses must be included because most particles in nature are massive. The usual way to do that relies on the validity of
 the CHY formula  in any number of dimension, so four-dimensional masses and off-shell momenta can  be
obtained from momenta in extra dimensions \cite{Nac14b,Nac15a,Nac15b}.
This works in some special cases, but not in general, because one cannot be assured that {\it all}  correct propagators can be obtained this way. 
For example, the propagators of a massive $\f^3$ theory cannot be so obtained 
if the total number of external particles far exceeds the total number of extra dimensions. This is further explained in Appendix A.

There is however a different method to generate a massive $\f^3$ theory and/or off-shell momenta that guarantees  
correct propagators
 in any number of dimension, achieved simply 
 by  modifying the scattering equations \cite{LY15a, DG19}. 
The method has also been used to extend the CHY Yang-Mills theory off-shell \cite{Lam19}. 

To reproduce the Standard Model, massive fermions in the fundamental representation must be included.  
That proves to be quite difficult in the CHY formalism although some interesting progress has been made \cite{Wein14b,HZ16a}.

Nature  contains many massive particles in a variety of interactions,  so the CHY formula 
must also be generalized to describe them, on-shell and  off-shell.
In this article we propose a method to do so for  scalar particles, again by  modifying the scattering equations.  
 We shall work out in detail the Yukawa coupling of several scalar nucleons and pions, all with
 different masses. In the special case of an infinite pion mass, this degenerates into the $\f^4$ theory.  We also discuss
 how to couple three different scalar fields via $\f_1\f_2\f_3$ interaction. In addition, a very useful  method to
 construct composite models from existing theories will be discussed. In this way we can for example construct a model in which
 two nucleons of different masses couple to form a di-nucleon resonance, and the resonances can interact
 with one other via an exchange of pions.  
Beyond scalar particles we also take a first look at photons. Compton scattering and bremsstrahlung from  a charged 
massive scalar particle can both be computed this way.

A disk function $Z$ and a sphere function $J$ have been introduced to connect massless field theory amplitudes to string amplitudes 
\cite{BSS13, CMS16, SS18}.
The modification of scattering functions mentioned above naturally leads to  a possible modification of the $Z$ and $J$ functions.
It would be interesting to see whether such a modification can be used to generate new string amplitudes and/or new effective field 
theories.

In Sec.~II, the general setup of the method is described,
together with illustrations taken from the two known cases:
massless on-shell $\f^3$ scattering \cite{CHY13c}, and massive on-shell or off-shell scattering  \cite{LY15a}.
This method is then applied to the Yukawa coupling of one massive scalar nucleon and one massive pion in
Sec.~III, and to the Yukawa coupling of several massive scalar nucleons and pions in Sec.~IV. 
Three field coupling $\f_1\f_2\f_3$ is taken up in Sec.~V, and the construction of composite models is
discussed in Sec.~VI.  Compton scattering and bremsstrahlung from charged scalar particles will be considered in Sec.~VII and Appendix C.
Modified $Z$ and $J$ functions will be discussed in Sec.~VIII, a short summary is included in Sec.~IX.
Appendix A explains why extra dimensions cannot be simply used to obtain massive
and/or off-shell amplitudes in four dimensions, and Appendix B proves a general covariant condition needed
for the CHY formula to be extended.

\section{The CHY $\boldsymbol{\f^3}$ Theory and Its Generalization}
\subsection{Equation of motion}
Consider a meromorphic function $F(\s)$ defined by
\be F(\s)=\sum_{i\not=j=1}^n{a_{ij}\over(\s-\s_i)(\s-\s_j)}=\sum_{i=1}^n{1\over \s-\s_i}\sum_{j\not=i, j=1}^n{a_{ij}\over\s-\s_j}:=\sum_{i=1}^n{\hat f_i(\s)\over\s-\s_i},\labels{F}\ee
where $\s_i,\  a_{ij}=a_{ji}$ are arbitrary complex parameters subject to the constraints $a_{ii}=0$, and 
\be \sum_{j=1}^na_{ij}=0, \quad (1\le i\le n).\labels{a}\ee
 Under a \M transformation,
\be \s&\to&{\a\s+\b\over\g\s+\d},\quad \s_i\to{\a\s_i+\b\over\g\s_i+\d},\quad \a\d-\b\g=1,\quad 1\le i\le n,
\labels{M}\ee
and as a consequence of the constraint, $\hat f_i(\s)$ can be shown to transform covariantly as
\be \hat f_i(\s)&\to&\l^2\hat f_i(\s),\quad{\rm where}\  \l=(\g\s+\d).\labels{cov}\ee 

The residue of $F(\s)$ at $\s=\s_i$ is
\be f_i=\sum_{j\not=i,j=1}^n{a_{ij}\over \s_i-\s_j}=\hat f_i(\s_i).\labels{f_i}\ee
Given an $a_{ij}$, $F(\s)=0$ for all $\s$ if and only all its residues are zero,  which is so
if and only if the $n$ parameters $\s_i$ are solutions of the $n$
{\it scattering equations} $f_i=0$. In light of covariance, three $\s_i$ can be chosen arbitrarily, so
only $n\-3$ of these $f_i$ can be linearly independent.
This linear dependence is encoded in the three sum rues 
\be \sum_{i=1}^n f_i&=&0,\nn\\
\sum_{i=1}^n f_i\s_i&=&0,\nn\\
\sum_{i=1}^n f_i\s_i^2&=&0.\labels{sr}\ee

Both \eq{cov} and \eq{sr} are known to be true for the original CHY massless on-shell $\f^3$ theory
\cite{CHY13a, CHY13b, CHY13c}. They are also true for the massive off-shell $\f^3$ theory \cite{LY15a}. That they are  
generally true for any $a_{ij}$ satisfying
\eq{a} is shown in Appendix B. It is this  covariant property that allows the CHY formula
\eq{chy} to be extended to cover many other field  theories.

We shall refer to $F(\s)=0$ as the {\it equation of motion} (EOM) because it generates all the scattering
equations, and because it resembles the Klein-Gordon equation
 in ordinary field theory. In a field theory, the EOM is a
differential equation true for all spacetime points $x$. In the CHY theory, it is an algebraic equation true
for all complex variables $\s$.

\subsection{CHY amplitudes}
Solutions of the scattering equations  determine the $n$-particle tree amplitude $A_n$. One way to
express that is through the CHY formula 
\be
A_n=\(-{1\over 2\pi i}\)^{n-3}\oint_\Gamma\s_{(rst)}^2\(\prod_{i=1,i\not=r,s,t}^n{d\s_i\over f_i}\){\cal I}_n,\labels{chy}\ee
where $\Gamma$ is a clockwise contour encircling all $f_i=0$, with $f_i$  given by \eq{f_i}, and ${\cal I}_n$  is chosen so that
 $A_n$ is invariant under the \M transformation
\eq{M}. \M invariance permits a choice of three arbitrary gauge constants
$\s_r, \s_s, \s_t$  in the integrand that does not affect the 
outcome of $A_n$ in  \eq{chy}.

Each $f_i$ factor plays the role of an inverse 
scalar propagator, and the ${\cal I}_n$
factor plays the role of interaction vertices. 
For scalar theories which we shall concentrate mostly on in this article, 
\be {\cal I}_n={1\over\s_{(12\cdots n)}\s_{(\a)}},\labels{In}\ee
where $\a=(\a_1\a_2\cdots\a_n)$ is a permutation of $(12\cdots n)$
determined by the specific Feynman diagram under consideration. The quantity $\s_{(p_1p_2\cdots p_k)}$
stands for the product $\prod_{i=1}^k\s_{p_ip_{i+1}}$, with $k+1\equiv 1$, and $\s_{ab}=\s_a-\s_b$. When spin-1 particles
are involved, ${\cal I}_n$ would be different. We shall discuss a special case in Sec.~VII.

It is implicitly assumed in \eq{In} that every Feynman tree diagram is written as a planar diagram, with the
external lines  ordered clockwise and cyclically in the natural order $1,2,\cdots, n$. The amplitude \eq{chy}
is that of a single Feynman diagram, or that of a sum of several or all Feynman diagrams, all depending on what
$\s_{(\a)}$ is. See Sec.~IIE for more discussions on that point.

The parameters $a_{ij}$ are so far arbitrary except for the constraint in \eq{a}. They will be chosen to 
reproduce the correct propagators  in a given theory. 
When  the external lines are ordered in the way described in the last paragraph, every propagator 
takes on the form $1/(k_S^2-m_S^2)$, where $k_S=\sum_{i\in S}k_i$ is the total external
momentum of a set $S$ of $\le (n-2)$ {\it consecutive} external lines, 
$k_i$ is the outgoing momentum
of line $i$, and $m_S^2$ is the (square)
mass of the propagator. In a CHY scalar theory given by \eq{chy},
the inverse propagators are
related to the parameters $a_{ij}$ by the formula \cite{LY15a}
\be k_S^2-m_S^2=\sum_{ i,j\in S}a_{ij}.\labels{pc}\ee
Different theory specifies different  $m_S^2$, and it is from this propagator equation that all  $a_{ij}$ are determined.
Once $a_{ij}$ are determined this way, we must still check that the symmetry condition $a_{ij}=a_{ji}$ and
the covariant condition \eq{a} are satisfied.
Otherwise \eq{chy} cannot be used. Moreover, we must make sure that \eq{pc} is self consistent in the following sense. 

For any set $S$ of consecutive external lines, let $\overline S$ be its complementary set, consisting of all the external lines not in $S$.
Since $S$ and $\overline S$ share the same propagator in a tree diagram, 
we must make sure that $k_S^2-m_S^2=k_{\overline S}^2-m_{\overline S}^2$.
With momentum conservation, $k_S=-k_{\overline S}$,  so this is satisfied if and only if
\be m_S^2=m_{\overline S}^2\labels{mcomp}\ee
for every set $S$. We shall refer to that as the {\it complementary condition}.

The two cases in which $a_{ij}$ are known are
the original CHY massless on-shell $\f^3$ theory \cite{CHY13c}, and
the massive off-shell $\f^3$ theory \cite{LY15a}. They will be reviewed in the following subsections. After that,  
we shall elaborate on how the propagator condition \eq{pc} can be used to determine $a_{ij}$ for other theories, and then 
proceed to apply the method to  a scalar
Yukawa theory whose `scalar nucleon' and `pion' possess different masses. 
In the limit of an infinite pion mass, this becomes the $\f^4$ theory. This Yukawa theory can be generalized to 
include many kinds of nucleons with different masses, provided they interact with one another only through
the exchange of pions.
In these theories, $a_{ij}$ determined by
the propagator condition can be made to obey the symmetry and the covariant conditions, as well as \eq{mcomp}. 
We will also discuss a theory with
three fields interacting according to $\f_1\f_2\f_3$. The propagator condition once again determine its
$a_{ij}$ for arbitrary masses of the three fields, but it turns out that such a $a_{ij}$ satisfies
the required conditions only when the three masses are identical. Composite models and photons, as well as modified
disk and sphere functions, will  be discussed in later sections.

\subsection{$\boldsymbol{\f^3}$ theory, on-shell and massless}
This is the original CHY theory \cite{CHY13c}. With $k_i^2=0$ and $m_S=0$, \eq{mcomp} is automatically satisfied.
The choice of $a_{ij}$ for $i\not=j$ is
\be a_{ij}=k_i\.k_j,\quad i\not=j. \labels{s0}\ee
The symmetry condition $a_{ij}=a_{ji}$ is obviously true. 
Owing to  momentum conservation and the massless requirement, the covariant requirement
\eq{a} is also satisfied because $\sum_{j=1}^na_{ij}=-k_i^2=0$.
The propagator condition \eq{pc}  with $m_S^2=0$ also follows because 
\be \sum_{ i,j\in S}a_{ij}=\sum_{i\in S}\sum_{j\in S}k_i\.k_j=k_S\.k_S=k_S^2.\nn\ee

\subsection{$\boldsymbol{\f^3}$ theory, off-shell and with mass $\boldsymbol{m}$}
This situation was studied in \cite{LY15a}. The solution with $a_{ii}=0$ is 
\be a_{ij}=k_i\.k_j+\r_{ij}-\h\m_{ij},\quad {\rm for\ }i\not=j.\labels{s1}\ee
The $\r_{ij}=\r_{ji}$ term is responsible for off-shell extensions, and the $\m_{ij}=\m_{ji}$ term 
is needed for a non-zero mass $m$.
Specifically, for $n>4$,
\be \r_{i,i\pm 1}&=&{1\over 2}(k_i^2+k_{i\pm 1}^2),\quad \m_{i,i\pm 1}= m^2,\nn\\
\r_{i,i\pm 2}&=&-{1\over 2}k_{i\pm 1}^2,\quad \m_{i,i\pm 2}=- m^2,\nn\\
\r_{ij}&=&0,\quad \m_{ij}=0, \quad {\rm for\ } |j-i|>2.\labels{s2}\ee
The indices $i\pm 1$ and $i\pm 2$ could turn out to be non-positive, or larger than $n$. In that case
they should be interpreted to have a value mod $n$.

For $n\le 4$, $j$ could be equal to or less than two lines away on {\it both} sides of $i$. In that case
\eq{s2} should be used to add up both sides to get the correct result for $\r_{ij}$ and $\m_{ij}$.

Since all the propagators have mass $m^2$, \eq{mcomp} is satisfied. With this $a_{ij}$, the covariant condition \eq{a} is satisfied because

\be \sum_{j=1}^na_{ij}&=&\sum_{j\not=i,j=1}^n(k_i\.k_j+\r_{ij}-\h\m_{ij})\nn\\
&=&-k_i^2+\r_{i,i+1}+\r_{i,i-1}+\r_{i,i+2}+\r_{i,i-2}-\h(\m_{i,i+1}+\m_{i,i-1}+\m_{i,i+2}+\m_{i,i-2})\nn\\
&=&-k_i^2+\h(k_i^2+k_{i+1}^2-m^2)+\h(k_i^2+k_{i-1}^2-m^2)-\h(k_{i+1}^2-m^2)-\h(k_{i-1}^2-m^2)\nn\\
&=&0.\labels{s3}\ee
To show that the propagator requirement
\eq{pc} is also satisfied, let $S=\{p, p+1, p+2, \cdots, q\}$, with $|S|:=q\-p\+1\le n\-2$. The latter
condition is necessary to ensure that there are at least two  lines in $S$ to merge to form
 a propagator somewhere.
Since $a_{ii}=0$,
\be \sum_{i,j\in S}a_{ij}&=&\sum_{i\not=j;i,j\in S}a_{ij}=2\sum_{i=p}^{q-1}\sum_{j=i+1}^{q}(k_i\.k_j+\r_{ij}-\h\m_{ij}).\labels{aij}\ee
For $j>i$, the only non-zero $\r_{ij}$'s are for $j=i+1$ and $j=i+2$. Hence
\be
2\sum_{i=p}^{q-1}\sum_{j=i+1}^{q}(k_i\.k_j+\r_{ij})&=&2\sum_{i=p}^{q-1}\sum_{j=i+1}^{q}k_i\.k_j+2\sum_{i=p}^{q-2}\[\r_{i,i+1}+\r_{i,i+2}\]-2\r_{q-2,+q-1}\nn\\
&=&2\sum_{i=p}^{q-1}\sum_{j=i+1}^{q}k_i\.k_j+\sum_{i=p}^{q-2}\[(k_i^2\+k_{i+1}^2)\-k_{i+1}^2\]\+
(k_{q-1}^2\+k_q^2)\nn\\
&=&(\sum_{i=p}^{q}k_i)^2=k_S^2,\labels{sar}\ee
and
\be
\sum_{i=p}^{q-1}\sum_{j=i+1}^{q}\m_{ij}&=&\sum_{i=q}^{q-2}\[\m_{i,i+1}+\m_{i,i+2}\]-\m_{q-2,q-1}=m^2.\labels{sm0}
\ee
Thus the propagator condition \eq{pc} is satisfied with $m_S^2=m^2$.

\subsection{Connection between $\boldsymbol{\s_{(\a)}}$ and Feynman diagrams}
Given a planar tree diagram such as Fig.1(a) whose external lines are ordered clockwise and cyclically according
to $(12\cdots n)$,  there are several techniques \cite{CHY13c,CG15,BBBD15a,BBBD15b,BBBDF15,LY15b,Lam18}
to determine the appropriate $\s_{(\a)}$  that yields the diagram. In this subsection we review how to do that  following the approach of \cite{LY15b,Lam18}.
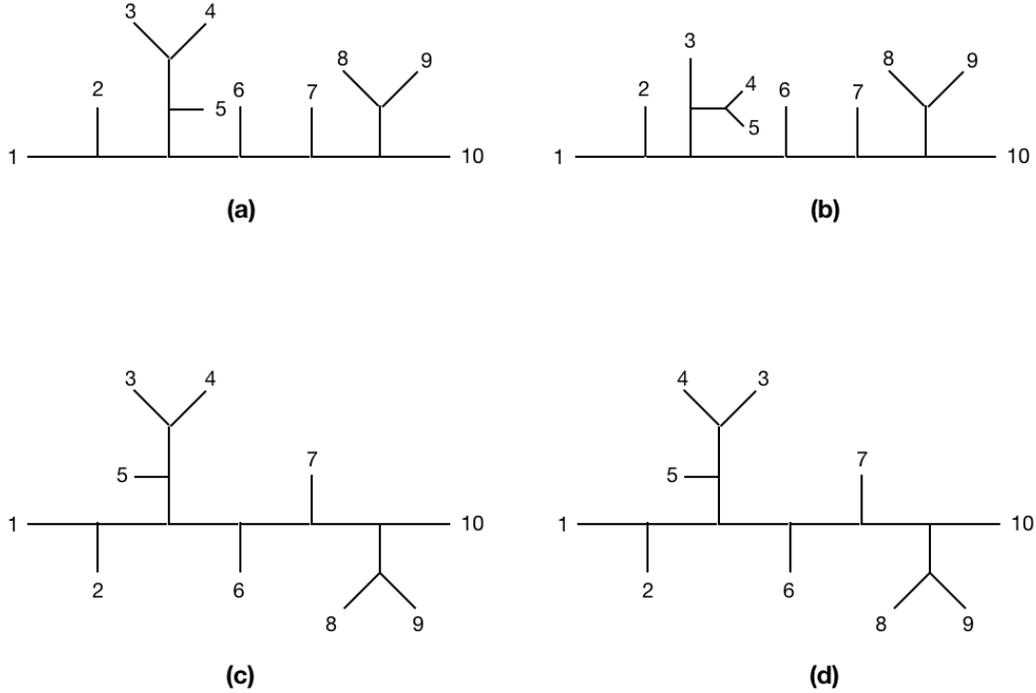
\bc\igw{14}{Fig1}\\ Fig.1\quad $\s_{(\a)}$ and propagators\ec

To that end, note that there are many ways to redraw a diagram 
by flipping its external lines. For Fig.1(a), one of them is 
Fig.1(c), another  one is Fig.1(d), and there are many others. 
The flipped diagrams possess the same  propagators 
as the original diagram, hence  the same scalar amplitude, but their external lines are cyclically ordered  
in  different ways. One of them would give the $\s_{(\a)}$ we are after.

To find out which, we have to know the general relation between propagators and $\s_{(\a)}$. 
For planar diagrams, propagators come from a merging of {\it consecutive} external lines.
It turns out that consecutive 
external lines labelled by $\a_u, \a_{u+1},\cdots,\a_{u+m}$ would merge into a propagator if and only if
they form a permutation of $m$ consecutive numbers $p, p\+1,\cdots, p\+m$. When that
happens we will put a square bracket around them, as in $[\a_u \a_{u+1}\cdots\a_{u+m}]$. A Feynman
tree diagram has $n\-3$ propagators, so $(\a)$ must contain $n\-3$ such {\it compatible} square brackets.
Two brackets are compatible if they either do not overlap, or one is completely
inside another. 

If there is only one way to partition  $(\a)$ into such compatible square brackets,
then this $\s_{(\a)}$ would give rise to one Feynman diagram. If there is more than one way to
partition $(\a)$, then the amplitude would receive contribution from several Feynman diagrams,
each corresponding to one such partition.

For example, the ordering  
$\a=(6215347(10)98)$ in Fig.1(c) can be partitioned only one way, into
$([6 [ [21] [5 [34] ] ] ] 7 [ (10) [98] ])$, so this is the right $\a$ for the single diagram Fig.1(a).
However, $\a=(6215437(10)98)$ in Fig.1(d) has two compatible partitions, 
$([6 [ [21] [5 [43] ] ] ] 7 [ (10) [98] ])$ and $([6 [ [21] [ [54] 3] ] ]  7 [ (10) [98] ])$, so that $\s_{(\a)}$  gives
rise to both Fig.1(a) and Fig.1(b), when the external lines are ordered according to $(12\cdots n)$.
More complicated $\a$ can give rise to more allowed partitions and more Feynman diagrams. In particular,
the identity permutation $\s_{(\a)}=\s_{(12\cdots n)}$  yields a sum of all Feynman tree diagrams.

In summary, given a single diagram, the appropriate $\s_{(\a)}$ is given by that flipping 
of external lines that produces a single  compatible partition for $(\a)$. 
A different flipping may result in a sum of several diagrams, and $\s_{(\a)}=\s_{(12\cdots n)}$
 would give rise to a sum of all Feynman diagrams.

This rule works not only for the $\f^3$ theory, but also all the other theories 
to be discussed in the following sections.

It would be useful to know  the origin of this rule \cite{LY15b,Lam18}, so that it can be generalized to
the situation  in Sec.~VII.  The $(n-3)$-fold integration in \eq{chy} is taken over every $\s_i$,  except
$\s_r, \s_s, \s_t$. Contribution to the integral comes from simple poles in the factor $1/\s_{(\a)}$, 
 located when the $\s_i$'s inside a square bracket coincide with one another. 
 When the numbers inside a square bracket are permutations of a consecutive subset of $(12\cdots n)$, a pole also occurs in the
 other factor $1/\s_{(12\cdots n)}$ of \eq{In}, but this other pole is compensated by a zero found in the factor $1/\prod_{i\not=r,s,t}f_i$,
 leaving behind a propagator as the residue. In this way the $(n-3)$ integrations in \eq{chy} give rise to $(n-3)$ propagators that
 make up a Feynman diagram.

\subsection{Beyond the ${\boldsymbol \f^3}$ theory}
If we
decompose $a_{ij}$ in the form of \eq{s1}, with $\r_{ij}$ given by \eq{s2}, then \eq{s3} and \eq{sar} show
that the symmetry condition $a_{ij}=a_{ji}$, covariant condition \eq{a},
 and the propagator condition \eq{pc} are already satisfied by 
the momentum part of $a_{ij}$.  That leaves only the mass part $\m_{ij}$,  which must  satisfy
\be
\m_{ij}&=&\m_{ji},\qquad ({\rm symmetry\ condition}),\labels{mcond0}\\
\sum_{j\not=i}\m_{ij}&=&0\qquad({\rm covariant\ condition}),\labels{mcond1}\\
\h\sum_{i\not=j, i,j\in S}\m_{ij}&=&m_S^2,\quad |S|\le n\-2,\qquad({\rm propagator\ condition}),\labels{mcond2}\ee
where $|S|$ is the number of lines in $S$.
Since $m_S^2$ depends on {\it all} the lines in $S$, $\m_{ij}$ must depend  on the particle nature of $i, j$, and all the lines in between. 
To exhibit such a dependence explicitly,
we shall denote $\m_{ij}$ by $(a_ia_{i+1}\cdots a_j)$ when $j>i$,
where $a$ is a particle identification symbol: $a_i$ specifies the particle of line $i$, $a_j$  the particle
of line $j$, and the remaining $a$'s specify the particle identity of lines in between. 
 The inequality $j>i$ means that line $j$ is downstream from
line $i$ when external lines are ordered cyclically in the clockwise direction.  It does not necessarily mean that the number $j$ is larger than the number $i$ because of the cyclic nature of the external lines.

Among the three terms of $a_{ij}$ in \eq{s1}, the $k_i\.k_j$ term is the simplest because it depends only
on lines $i$ and $j$. The $\r_{ij}$ term is more complicated because it can depend on the lines
between $i$ and $j$, {\it e.g.},  $\r_{i, i\pm 2}=\h k_{i\pm 1}^2$, but it becomes zero when $j$ and $i$
are more than two lines apart. The last term $\m_{ij}$ is the most complicated
because not only it depends on  all lines between $i$ and $j$,  it may also be non-zero no matter how
far apart lines $i$ and $j$ are. Moreover, it also depends on the precise ordering and identity of the 
particles in between. 

Since kinematics are universal, the $k_i\.k_j$ and $\r_{ij}$ terms are the same
in all theories. Thus it is the $\m_{ij}$ term that tells theories apart. 
However, even for the same theory, different ordered sets of external lines may give rise to different $\m_{ij}$'s.
 For example, in the Yukawa theory to be discussed
in the next section, where solid and dotted lines represent a scalar nucleons and  pions respectively,
the amplitude in Fig.2(a) and the amplitude in Fig.2(b) may have different sets of $\m_{ij}$. Diagrams (a1), (a2), (a3) are the tree diagrams contained
in Fig.2(a), and (b1), (b2), (b3), (b4) are the tree diagrams contained in Fig.2(b). $\m_{ij}$ for (a1), (a2), (a3) are the same, but that
may not be the same as the $\m_{ij}$ for (b1), (b2), (b3), and (b4).

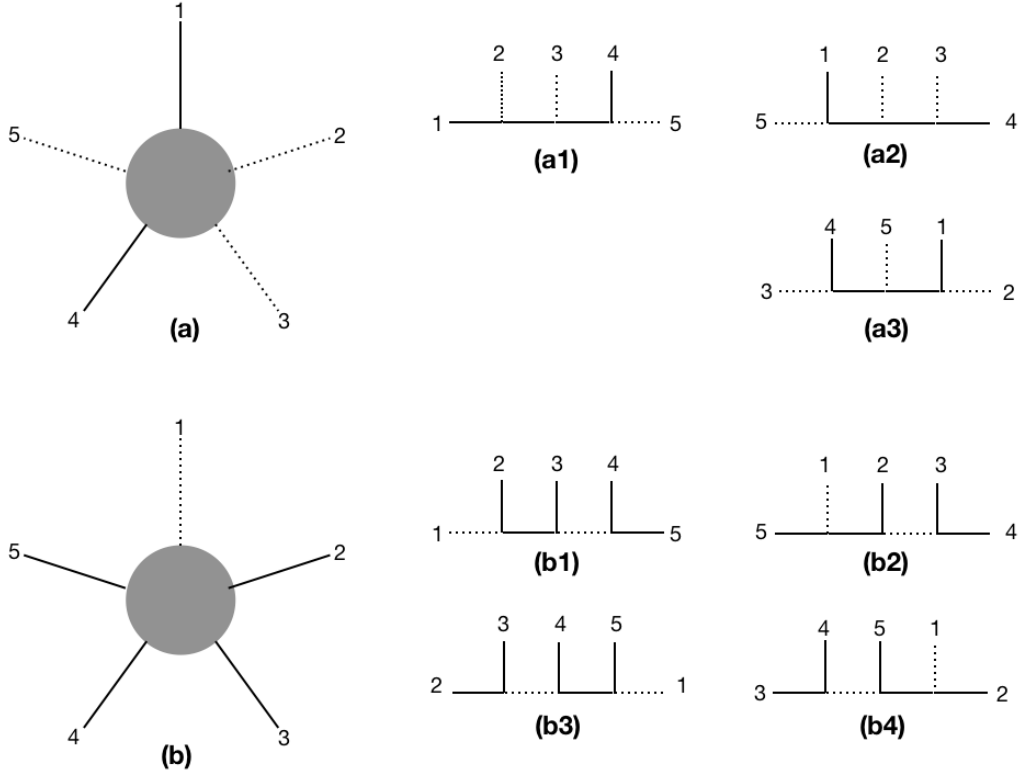
\bc\igw{14}{Fig2}\\ Fig.2\quad Two tree amplitudes for Yukawa interaction. Solid lines are nucleons, and dotted lines are pions\ec

Let us delve  a bit more into the details on how \eq{pc} can be used to compute $\m_{ij}$.
If $S$ consists of consecutive lines from $p$
to $q$, then we will write the left hand side of \eq{mcond2} as $\bk{a_pa_{p+1}\cdots a_{q}}$. With that
notation, \eq{mcond2} can be written as
\be
m_S^2=\bk{a_pa_{p+1}\cdots a_{q}}=\sum_{i=p}^{q-1}\sum_{j=i+1}^q\m_{ij}=
\sum_{i=p}^{q-1}\sum_{j=i+1}^q(a_ia_{i+1}a_{i+2}\cdots a_j).\labels{mcond3}\ee
This allows $\m_{ij}$ to be computed by four combination of $m_S^2$  using the formula
\be
\m_{ij}=(a_ia_{i+1}\cdots a_{j-1}a_j)&=&\bk{a_ia_{i+1}\cdots a_{j-1}a_j}-\bk{a_{i+1}\cdots a_{j-1}a_j}\nn\\
&-&\bk{a_ia_{i+1}\cdots a_{j-1}}+\bk{a_{i+1}\cdots a_{j-1}},\quad (j>i),\labels{mcond4}\ee
if $n-2\ge |S|=q-p+1\ge 4$. 
For smaller $|S|$, the appropriate formulas are
\be
(a_pa_{p+1})=\bk{a_pa_{p+1}},\quad (a_pa_{p+1}a_{p+2})=\bk{a_pa_{p+1}a_{p+2}}-\bk{a_pa_{p+1}}-\bk{a_{p+1}a_{p+2}}.\labels{mcond5}\ee
To use these formulas to determine $\m_{ij}$, we must know what $m_S^2$ is. That depends on  the propagator,
which in  turn depends on the nature of interaction and the particle content of the external lines in $S$. 
More specifically, an interaction determines a set of `topological requirements', which in turn fixes $m_S^2$.
It will become clear
that $m_S^2$ depends only on the particle content of $S$, but never on their ordering,  
so $\bk{a_pa_{p+1}\cdots a_q}$ is permutation invariant in the symbols. However,  $(a_pa_{p+1}\cdots a_q)$ {\it does} depend on how the lines are ordered. Nevertheless,
it follows from \eq{mcond4} and \eq{mcond5}  that the reversal relation
\be (a_pa_{p+1}\cdots a_{q-1}a_q)=(a_qa_{q-1}\cdots a_{p+1}a_p)\labels{mcond5aa}\ee
is always valid.

\subsection{Complementary, symmetry, and covariant conditions}
It will be shown in this subsection that once the complementary condition \eq{mcomp} is obeyed, the symmetry condition \eq{mcond0} and
the covariant condition \eq{mcond1} are automatically satisfied. For that reason, the important thing to check for each theory is whether the
complementary condition is fulfilled. In using the propagator condition to determine $\m_{ij}$, it turns out that often unknown parameters
have to be introduced. Forcing the complementary condition to be valid would then determine some or all of these parameters.
\vs
\noindent\underline{Symmetry condition}.\\ 
If $\m_{ij}$ is computed using \eq{mcond4} for $j>i$, then $\m_{ji}$ should be computed using \eq{mcond4} for $i>j$. Note that these two
inequalities are not mutually contradictory because they simply mean that the second index in each case is downstream from the first index
in a clockwise direction. It says nothing about the numerical sizes of $i$ vs $j$. With that in mind, we have 
\be
\m_{ij}&=&\bk{i,i\+1,\cdots, j\-1,j}-\bk{i\+1,\cdots, j\-1,j}-\bk{i,i\+1,\cdots, j\-1}+\bk{i\+1,\cdots, j\-1},\nn\\
\m_{ji}&=&\bk{j,j\+1,\cdots, i\-1,i}-\bk{j\+1,\cdots, i\-1,i}-\bk{j,j\+1,\cdots, i\-1}+\bk{j\+1,\cdots, i\-1}.\nn\\
&&\labels{mijji}\ee
Each of the four terms in $\m_{ij}$ contains lines that are complementary to lines in one term of $\m_{ji}$.
For example, the complement of the first term in $\m_{ij}$
is the last term of $\m_{ji}$. Using the complementary condition $m_S^2=m_{\overline S}^2$, it follows that  $\m_{ij}=\m_{ji}$. 
\vs

\noindent\underline{Covariant condition}\\
To how the covariant condition \eq{mcond1}  from the propagator condition,
it is important to note that \eq{mcond2} is valid only for $|S|\le n-2$. With that caveat and using the symmetry
condition for the last two terms,
\be
\sum_{j\not=i}\m_{ij}&=&\sum_{j=i+1}^{i-3}\m_{ij}+\m_{i-2,i}+\m_{i-1,i}\nn\\
&=&\bk{i,i+1,\cdots,i-3}-\bk{i+1,\cdots,i-3}+\bk{i-2,i-1,i}-\bk{i-2,i-1}.\label{covy}\ee
The lines in the first and last terms are complementary,  and the lines
in the second and third terms are also complementary. Thus \eq{mcomp} ensures the right hand side of the above equation to vanish.

\subsection{Uniqueness of $\boldsymbol{a_{ij}}$}
$a_{ij}$ given by \eq{s1}, \eq{mcond2} and \eq{mcond4} is unique if we use these expressions on
 all possible propagators in all  Feynman diagrams of a given ordering of external lines. If we only want
to find an $a_{ij}$ that yields a single diagram, then there are many ways to choose it, because the propagator
condition \eq{pc} has to be satisfied only by $n\-3$ sets $S$, instead of all the consecutive sets. 

The same remark holds for all the other theories to be considered in later sections.

\section{Yukawa Theory and the $\boldsymbol{\f^4}$ theory}

\subsection{Topological and propagator conditions}
Consider  a scalar nucleon field $\f$ with  mass $m_1$
 and a pion field $\f'$ with  mass $m_0$, coupled via  a Yukawa interaction
 $\f^2\f'$. Each vertex has two nucleon lines and one
pion line; an $n$-point tree amplitude has $n-3$ propagators.
If $E, E'$ is the number of external lines  of nucleons and pions, and $I, I'$
is their number of internal lines, then
\be 
E+2I&=&2(E'+2I'),\nn\\
E+E'&=&n=I+I'+3.\labels{tc1}\ee
Solving for $I$ and $I'$, the solution is
\be I&=&\h E+E'-2,\nn\\
I'&=&\h E-1.\labels{tc2}\ee
In order for $I, I'$ to be non-negative integers,  $E$ must be an even integer $\ge 2$. When $E=2$, then we must have $E'\ge 1$.
This {\it topological requirement} applies not only to the whole Feynman diagram, it also applies to a sub-diagram consisting of a
propagator and all its corresponding external lines.
It is the basis from which $\m_{ij}$ for the Yukawa theory is determined.

If the set $S$ of consecutive external line consists of an even number of nucleons, then the propagator line must be a pion
so $m_S^2=m_0^2$. If the set consists of an odd number of nucleons, then the propagator line must
be a nucleon so $m_S^2=m_1^2$. If the set consists of only pions, then no propagator can be formed,
so $m_S^2$ is undetermined. We shall assign it a value $M^2$ where $M$ is so far arbitrary. To summarize,
using $N(S)$ to denote the number of nucleons in the set $S$, then
\be m_S^2&=&m_0^2,\qquad (N(S)={\rm even}\not=0),\nn\\
&=&M^2,\qquad (N(S)=0),\nn\\
&=&m_1^2,\qquad (N(S)={\rm odd}).\labels{mcond5a}\ee

\subsection{Complementary condition}
In order to satisfy the complementary condition $m^2_S=m^2_{\overline S}$  of \eq{mcomp}, we need to set $M^2=m_0^2$. 
This is so because the nucleon number of the whole diagram is even, so $(-)^{N(S)}=(-)^{N(\overline S)}$. In particular, if $N(S)$ is even, 
then it is guaranteed that
$N(\overline S)$ is even, but depending on the diagram and the configuration of external lines, it may or may not be zero. So in order for \eq{mcomp} to be 
satisfied for all conceivable sets $S$, we must let $M^2=m_0^2$ in \eq{mcond5a}.

\subsection{Illustrative examples}
Let $a=0$ denote a pion and $a=1$ a nucleon. 
Using \eq{mcond4}, \eq{mcond5}, and  \eq{mcond5a} with $M^2=m_0^2$, we get 
\be (00)&=&\{00\}=m_0^2,\quad (01)=(10)=\{01\}=m_1^2,\quad (11)=\{11\}=m_0^2,\nn\\
 (011)&=&(110)=\{110\}-\{01\}-\{11\}=-m_1^2,\quad (000)=\{000\}-2\{00\}=-m_0^2,\nn\\
  (100)&=&(001)=\{100\}-\{10\}-\{00\}=-m_0^2,\quad (101)=\{101\}-\{10\}-\{01\}=m_0^2-2m_1^2,\nn\\
   (111)&=&\{111\}-2\{11\}=m_1^2-2m_0^2.\labels{y23}\ee		
For $|S|\ge 4$,  $\m_{ij}$ can be written in the form
\be
(aVb)=\bk{aVb}-\bk{aV}-\bk{Vb}+\bk{V},\nn\ee
where $a, b,$ and the members of the subset $V$ are 0's or 1's. Using \eq{mcond5a}, it follows that $(aVb)=0$ if either $a$ or $b$ is 0, because the four terms cancel pairwise. 
The only non-zero $\m_{ij}$ is of the form $(1V1)$, and it is equal to $2(m_0^2-m_1^2)$ or
$2(m_1^2-m_0^2)$ depending on whether $V$ contains an even number of nucleons or an odd number of nucleons.

 These results are summarized in the following equation, where $a, b$ are either 0 or 1,
 and $N(V)$ stands for the number
 of nucleons in the subset $V$:
 \be
(00)&=&(11)=m_0^2,\quad (01)=(10)=m_1^2,\quad (011)=(110)=-m_1^2,\nn\\
(000)&=&(100)=(001)=-m_0^2,\quad (101)=m_0^2-2m_1^2,\quad (111)=m_1^2-2m_0^2,\nn\\
(0Vb)&=&(aV0)=0,\quad (|V|\ge 2),\nn\\
(1V1)&=&2(m_0^2-m_1^2)\ \ (N(V)={\rm even}),\quad (1V1)=2(m_1^2-m_0^2)\ \ (N(V)={\rm odd}).
\labels{yukawa}\ee

\subsubsection{Elastic nucleon-nucleon and pion-nucleon scatterings}

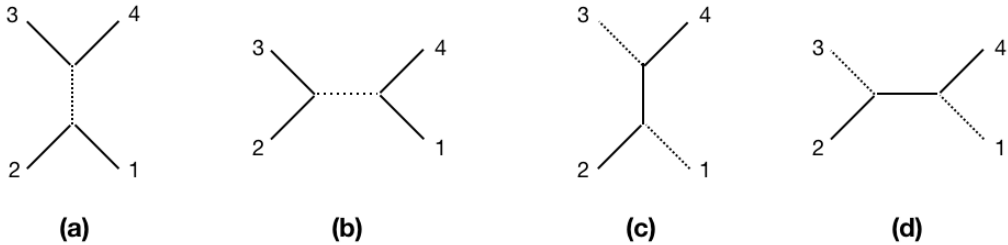
\bc\igw{14}{Fig3}\\ Fig.3\quad Elastic nucleon-nucleon and pion-nucleon scattering diagrams.
Solid lines are nucleons and dotted lines are pions\ec
Let us apply \eq{chy} to the scattering diagrams shown in Fig.3. By choosing  the \M constant lines $r, s, t$ to be $2, 3, 4$,
only $f_1$ and $\m_{1j}$ enter into \eq{chy}.

Recall that \eq{mcond3} can be used to compute $\m_{ij}$ only when $j-i+1\le n-2$, because it requires at least two
external lines to merge into a propagator. Thus for $n=4$,
\eq{yukawa} can be used  only when $j=i+1$. In particular, for $i=1$, only $\m_{12}$ can be so calculated, though
$\m_{14}$ can also be calculated using the symmetry condition $\m_{14}=\m_{41}=\m_{45}$. That leaves $\m_{13}$, which cannot be calculated from \eq{mcond3}, but it can be calculated
from \eq{mcond1} as $\m_{13}=-(\m_{12}+\m_{14})$.

 For Fig.3(a) and Fig.3(b), using \eq{yukawa},
\be
\m_{12}&=&(11)=m_0^2, \quad \mu_{14}=\m_{41}=(11)=m_0^2,\quad  \m_{13}=-(\m_{12}+\m_{14})=-2m_0^2.\labels{ysa}\ee

For Fig.3(c) and Fig.3(d),
\be
\m_{12}&=&(01)=m_1^2,\quad \m_{14}=\m_{41}=(10)=m_1^2,\quad m_{13}=-(\m_{12}+\m_{14})=-2m_1^2.\labels{ysb}\ee

The $\s_{(\a)}$ factor is $\s_{(2134)}$ for Fig.1(a) and Fig.1(c), and is $\s_{(4132)}$ for Fig.1(b) and Fig.1(d).
Thus the inverse propagators are 
$2a_{12}=2k_1\.k_2+k_1^2+k_2^2-\m_{12}=(k_1+k_2)^2-m_0^2$ for Fig.3(a),
$2a_{14}=2k_1\.k_4+k_1^2+k_4^2-\m_{14}=(k_1+k_4)^2-m_0^2$ for Fig.3(b),
$2a_{12}=2k_1\.k_2+k_1^2+k_2^2-\m_{12}=(k_1+k_2)^2-m_1^2$ for Fig.3(c), and
$2a_{14}=2k_1\.k_4+k_1^2+k_4^2-\m_{14}=(k_1+k_4)^2-m_1^2$ for Fig.3(d), all agreeing with what the diagrams indicate.

 Note  that $\m_{13}, \m_{14}$ are not needed for Figs.3(a) and 3(c), and $\m_{13}, \m_{12}$ are not needed for Figs.3(b) and 3(d),
so they could have been something else without affecting the propagator. See Sec.~IIH for more discussions on this point.

Note also that we can get Fig.1(a) and Fig.1(b) together, also Fig.1(c) plus Fig.1(d) at the same time, by choosing $\s_{(\a)}=\s_{(1234)}$.

\subsubsection{$\boldsymbol{N+N\to N+N+\pi+\pi}$}
There are many diagrams for the emission of two pions. For the purpose of illustration we
consider only Fig.4. In this case, by fixing the \M constant lines
$r, s, t$ to be 2, 5, 6, only $f_1, f_3, f_4$ appear in \eq{chy}. The corresponding mass terms calculated from \eq{yukawa} are
\be
\m_{12}&=&(11)=m_0^2,\quad \m_{13}=(111)=-2m_0^2+m_1^2,\quad \m_{14}=(1110)=0,\nn\\
 \m_{15}&=&\m_{51}=(011)=-m_1^2,\quad \m_{16}=\m_{61}=(11)=m_0^2.\nn\\
\m_{31}&=&\m_{13}=-2m_0^2+m_1^2,\quad \m_{32}=\m_{23}=(11)=m_0^2,\quad \m_{34}=(10)=m_1^2,\nn\\
\m_{35}&=&(100)=-m_0^2,\quad \m_{36}=(1001)=2(m_0^2-m_1^2).\nn\\
\m_{41}&=&\m_{14}=0,\quad \m_{42}=\m_{24}=(110)=-m_1^2, \quad \m_{43}=\m_{34}=m_1^2,\nn\\
\m_{45}&=&(00)=m_0^2,\quad \m_{46}=(001)=-m_0^2,\nn\\
\s_{(\a)}&=&\s_{(215346)}.\labels{ysd}\ee
It can be directly checked from these relations that the covariant requirement $\sum_{j\not=i}\m_{ij}=0$ is satisfied for $i=1, 3, 4$.

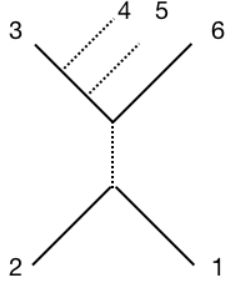
\bc\igw{4}{Fig4}\\ Fig.4\quad A nucleon-nucleon scattering diagram with the emission of two pions\ec

The three inverse propagators 
\be
2a_{12}&=&(k_1+k_2)^2-\m_{12}=(k_1+k_2)^2-m_0^2,\nn\\
2a_{34}&=&(k_3+k_4)^2-\m_{34}=(k_3+k_4)^2-m_1^2,\nn\\
2(a_{34}+a_{35}+a_{45})&=&(k_3+k_4+k_5)^2-(\m_{34}+\m_{35}+\m_{45})=(k_3+k_4+k_5)^2-m_1^2,
\ee
are as shown in the diagram. 

\subsection{$\boldsymbol{\s_{(\a)}}$ and sum over diagrams}
So far we have discussed how to choose $\s_{(\a)}$ and $f_i$ in  \eq{chy} 
to compute the amplitude of a single tree diagram in the Yukawa theory. 
For the $\f^3$ theory, Sec.~IIE shows how $\s_{(\a)}$ can also be chosen  
to yield  a
sum of several, or all, Feynman diagrams in the amplitude. 
Since the method of choosing $\s_{(\a)}$ has nothing to do with $f_i$, it is expected to
 be valid  for the Yukawa theory as well. This is indeed so, but with one caveat.   
 
 Recall the method  consists of partitioning $(\a)$ into $n-3$ compatible square brackets, each representing
 a propagator in a Feynman diagram. For  the Yukawa
 theory, this is still true as long as we avoid those  partitions containing square
 brackets with only pion lines. We must exclude those because pions alone cannot merge into a
 propagator, which is why we assign an arbitrary constant $M^2$ to that case in \eq{mcond5a}. If it were
 not for the complementary condition which forces $M^2$ to be $m_0^2$, even those diagrams
 need not be avoided because we could have set $M^2=\infty$ to get rid of them at the end. As it is,
 $M^2=m_0^2$, but this is {\it not} the propagator mass of several pions, because such a set does not even produce a propagator.
 To get things right, we must eliminate those phantom propagators by hand, by excluding the partitions containing pure pion square brackets. As long as that is followed, everything else
 is the same as in the $\f^3$ theory.
 
 For example, the particles in the six external lines of Fig.4 are of the type (111001), with lines 4 and 5 being
 two consecutive pion lines. By
  setting $\s_{(\a)}=\s_{(123456)}$, one would have included diagrams containing a two-pion phantom propagator and
  produced a wrong result, unless we exclude by hand partitions that have pure pion square brackets. 
  However, if the fifth
pion line were attached to the downstream side of the next nucleon, so that the particle content becomes (111010), then there are
no consecutive pion set so $\s_{(\a)}=\s_{(123456)}$ for that kind of $\m_{ij}$ would produce the correct sum of all Feynman tree diagrams
with those particle contents.

\subsection{The $\boldsymbol{\f^3}$ limit}
By setting $m_0=m_1=m$ we recover the $\f^3$ theory from the Yukawa theory. In that limit, it follows
from \eq{yukawa} that $\m_{i,i\pm 1}=m^2,\ \m_{i, i\pm2}=-m^2$, and $\m_{ij}=0$ for
$|j-i|>2$, which is what it should be according to \eq{s2}.

\subsection{The $\boldsymbol{\f^4}$ theory}
In the $m_0\to\infty$ limit, a multi-nucleon tree diagram such as Fig.5(a) becomes 
 a $\f^4$ diagram such as Fig.5(b), provided the coupling constant is suitably scaled by a power of $m_0$.
 However, it is important to remember that the $m_0\to\infty$ limit should be taken after the integral \eq{chy}
 is carried out.
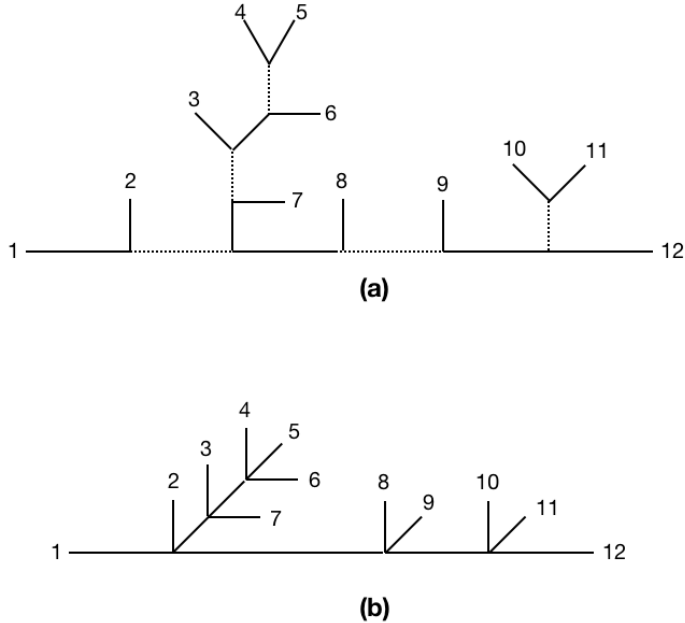
\bc\igw{10}{Fig5}\\ Fig.5\quad A diagram with $\f^4$ vertices such as (b) can be obtained from a Yukawa 
diagram such as (a) by setting the pion mass to be infinity\ec

\section{Multiple Particle Yukawa Theory}
The Yukawa theory of last section can be generalized to include several kinds of scalar nucleons and pions, all
with different masses. 
In order not to make the model and the notation  too complicated, we consider here only two kinds of nucleons, 
with masses $m_1$ and $m_2$ respectively. Each kind can interact with
pions of mass $m_0$ via the Yukawa interaction $\f_1^2\f'+\f_2^2\f'$. For simplicity, we have also
set both coupling constants equal to 1.

\subsection{Propagator condition}
There must be an even number of external nucleons of each kind in a Feynman diagram,
or a sub-diagram with one propagator. Moreover, in the tree 
approximation, there can be no Feynman diagram with external pions alone. 
Let $N_1, N_2$ be the number of external lines for the two kinds of nucleons,
then $m_S^2$ for a set $S$ of consecutive external lines can be determined as before to be
\be
m_S^2&=&m_0^2\quad {\rm if\ } (N_1, N_2)={\rm(even,\ even)}\not=(0,0),\nn\\
&=&M^2 \quad {\rm if\ } (N_1, N_2)=(0,0),\nn\\
&=&m_1^2\quad  {\rm if\ } (N_1, N_2)={\rm(odd,\ even)},\nn\\
&=&m_2^2\quad  {\rm if\ } (N_1, N_2)={\rm(even,\ odd)},\nn\\
&=&\overline M^2\quad  {\rm if\ } (N_1, N_2)={\rm(odd,\ odd)}.\labels{twoyu}\ee
Both $M^2, \overline M^2$ are arbitrary parameters because these sets of external lines cannot merge into
a propagator. 

\subsection{Complementary condition}
Using the fact that the whole diagram must of an even number of nucleons of each kind, one can conclude that 
$(-)^{N_1(S)}=(-)^{N_1(\overline S)}$ and 
$(-)^{N_2(S)}=(-)^{N_2(\overline S)}$. From \eq{twoyu}, it follows that 
the complementary condition \eq{mcomp} is
satisfied if and only if $M^2=m_0^2$. Note that the unknown parameter ${\overline M}^2$ remains undetermined, so it must not enter
into any propagator in any diagram. 

\subsection{Illustrative example}
As before, the propagator condition \eq{mcond3} can be used to compute $\m_{ij}$.
Here are some results for small $|S|$. The reversal condition \eq{mcond5aa} can get us some more. Let $a=0, 1, 2$ denote a pion, the first kind of nucleon, and the second
kind of nucleon.  By setting $M^2=m_0^2$ and using  \eq{mcond4} and \eq{mcond5}, one gets, for $|S|=2$,
\be
(00)&=&m_0^2, \quad (12)=\Mb, \quad (11)=(22)=m_0^2,\quad (01)=m_1^2,\quad
(02)=m_2^2.\labels{twoyu0}\ee
For $|S|=3$,
\be
(000)&=&-m_0^2, \quad (111)=\bk{111}-\bk{11}-\bk{11}=m_1^2-2m_0^2,\quad 
(222)=m_2^2-2m_0^2,\nn\\
 (012)&=&\bk{012}-\bk{01}-\bk{12}=-m_1^2,\quad (021)=\bk{021}-\bk{02}-\bk{21}=-m_2^2,\nn\\
  (011)&=&\bk{011}-\bk{01}-\bk{11}=-m_1^2,\quad (022)=\bk{022}-\bk{02}-\bk{22}=-m_2^2,\nn\\
 (102)&=&\bk{102}-\bk{10}-\bk{02}=\Mb-m_1^2-m_2^2,\quad (101)=\bk{101}-\bk{10}-\bk{01}=m_0^2-2m_1^2,\nn\\
  (202)&=&\bk{202}-\bk{20}-\bk{02}=m_0^2-2m_2^2,\quad  (001)=\bk{001}-\bk{00}-\bk{01}=-m_0^2,\nn\\
 (002)&=&\bk{002}-\bk{00}-\bk{02}=-m_0^2,  \quad (010)=\bk{010}-\bk{01}-\bk{10}=-m_1^2,\nn\\
 (020)&=&-m_2^2,\quad (112)= \bk{112}-\bk{11}-\bk{12}=m_2^2-m_0^2-\Mb,\nn\\
 (221)&=&m_1^2-m_0^2-\Mb, \quad (121)=\bk{121}-\bk{12}-\bk{21}=m_2^2-2\Mb,\nn\\
 (212)&=&m_1^2-2\Mb.\labels{twoyu1}
\ee
For $|S|=4$, only a few results are listed below for illustration, but clearly we can compute all the results
for every $|S|$.
\be
(1120)&=&\bk{1120}-\bk{112}-\bk{120}+\bk{12}=0,\nn\\
(1202)&=&\bk{1202}-\bk{120}-\bk{202}+\bk{20}=m_1^2-\Mb-m_0^2+m_2^2,\nn\\
(2020)&=&\bk{2020}-\bk{202}-\bk{020}+\bk{02}=0,\nn\\
(0201)&=&\bk{0201}-\bk{020}-\bk{201}+\bk{20}=0.\labels{twoyu2}
\ee

By taking the \M constant lines $r, s, t$ to be lines 4, 5, 6 in Fig.6, only $f_1, f_2, f_3$ enter into \eq{chy}.
The relevant $\m_{ij}$ can be computed from \eq{twoyu1}, \eq{twoyu2}  to be
\be
\m_{12}&=&(11)=m_0^2,\quad \m_{13}=(112)=m_2^2-m_0^2-\Mb, \quad \m_{14}=(1120)=0,\nn\\
\m_{15}&=&\m_{51}=(201)=\Mb-m_1^2-m_2^2,\quad \m_{16}=\m_{61}=(01)=m_1^2.\nn\\
\m_{21}&=&\m_{12}=(11)=m_0^2,\quad \m_{23}=(12)=\Mb, \quad \m_{24}=(120)=-m_2^2,\nn\\
\m_{25}&=&=(1202)=m_1^2-\Mb-m_0^2+m_2^2,\quad \m_{26}=\m_{62}=(011)=-m_1^2.\nn\\
\m_{31}&=&\m_{13}=(112)=m_2^2-m_0^2-\Mb,\quad \m_{32}=\m_{23}=\Mb, \quad \m_{34}=(20)=m_2^2,\nn\\
\m_{35}&=&(202)=m_0^2-2m_2^2,\quad \m_{36}=(2020)=0.\labels{twoyumij}\ee
Using these formulas, it can be explicitly checked that $\sum_{j\not=i}\m_{ij}=0$ for $i=1, 2, 3$. 
The inverse propagators in Fig.6  can be obtained from \eq{pc}, \eq{s1}, and \eq{twoyumij} to be
\be
(k_6+k_1+k_2)^2-(\m_{16}+\m_{26}+\m_{12})&=&(k_6+k_1+k_2)^2-m_0^2,\nn\\
(k_6+k_1)^2-\m_{16}&=&(k_6+k_1)^2-m_1^2,\nn\\
(k_3+k_4)^2-\m_{34}&=&(k_3+k_4)^2-m_2^2.\nn\ee

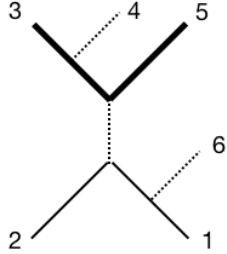
\bc\igw{4}{Fig6}\\ Fig.6\quad Dotted lines are pions, and thin (thick) solid lines are nucleons 1 (2).\ec

Note that
although the unknown parameter $\Mb$ may be contained in individual $\m_{ij}$, they drop out in the
covariant sum, and they also drop out in the propagator masses. For example, the propagator
in the second nucleon
can also be obtained also from lines 5, 6, 1, 2, so we should also have $m_2^2=\m_{12}+\m_{15}+\m_{16}
+\m_{25}+\m_{26}+\m_{56}$. Except for $\m_{56}$, which is $(20)=m_2^2$, all the other $\m_{ij}$ already
appear in \eq{twoyumij}. One can check although  $\Mb$ appears in $\m_{15}$ and $\m_{25}$,
it disappears from the sum, and the sum  does add up to $m_2^2$.

\section{$\boldsymbol{\f_1\f_2\f_3}$ Theory}

\subsection{Topological rules for tree amplitudes}
For easy reference, we shall refer to the three  fields as having  three different `colors'.
Consider a connected tree diagram with $E_a\  (I_a)$ external (internal) lines of color $a$, and a total of $n$
external lines.  Such a diagram has $n-3$ internal lines, and $n-2$
vertices, hence
\be
E_1+E_2+E_3&=&n=I_1+I_2+I_3+3,\nn\\
E_1+2I_1&=&n-2=E_2+2I_2=E_3+2I_3.\labels{tr1}\ee
Given $E_a$, these rules lead to the solution
\be
I_1&=&{1\over 2}(E_2+E_3)-1,\nn\\
I_2&=&{1\over 2}(E_1+E_3)-1,\nn\\
I_3&=&{1\over 2}(E_2+E_1)-1.\labels{tr2}\ee
In order for $I_a$ to be a non-negative integer, we must have $E_1, E_2, E_3$ to be either all odd,
or all even numbers. We must also have $E_a+E_b\ge 2$ for every pair $a, b$, thereby forbidding
diagrams with only one color of external lines.
These {\it topological
requirements} are important for the determination of $\m_{ij}$.

Fig.7 gives a concrete illustration of \eq{tr2}, with
$n=10$, $[E_1,E_2,E_3]=[4,4,2]$, and $[I_1,I_2,I_3]=[2,2,3]$.
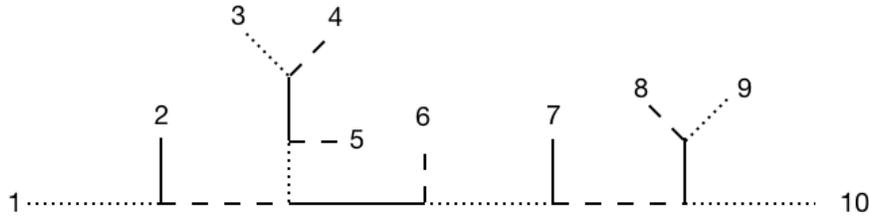
\bc\igw{12}{Fig7}\\ Fig.7\quad A $\f_1\f_2\f_3$ tree diagram, with $\f_1, \f_2, \f_3$ depicted respectively by 
dotted, dashed, and solid lines. \ec

Using the  topological requirements,  a set $S$ of consecutive lines can be sorted into 10 different classes.
$T_0, T_1, T_2, T_3$ are classes where the external lines in $S$ cannot merge into a propagator,
and $S_1, S_2, S_3, S'_1, S'_2, S'_3$ are classes that can. These classes are distinguished by the
number of external lines $E_a^S$ of color $a$ in the class.  The resulting sub-diagram, consisting of 
the propagator and the lines in $S$, must itself satisfy the topological requirement.  This requirement will
be used to determine the propagator mass $m_S^2$.

$T_0$ is the class where
$E^S_a$ are all even or all odd. $T_a\ (a=1, 2, 3)$ is the class where all lines have the same color $a$.
None of these classes can merge into a propagator, for otherwise the topological requirement would be
violated for the sub-diagram whatever color  the propagator carries.  With no propagator, $m_S^2$  is undetermined. 
We shall assign it to be
 $M_0^2$ for class $T_0$, and $M_a^2$ for class $T_a$. They may remain arbitrary, in which case they
 should not appear in the final expression of $A_n$ given by \eq{chy}, or they may be determined by the complementary
condition.
 
 For illustration, here are some concrete examples of the different classes.  
$[E^S_1, E^S_2, E^S_3]=[0,5,5]$ belongs to
class $S_1$, [0, 2, 5] belongs to class $S_3'$, and [0, 6, 0] belongs to class $T_2$. 
The diagram Fig.7 as a whole
has $[E^S_1, E^S_2, E^S_3]=[4, 4, 2]$, so it belongs to class $T_0$.

 If $[E^S_1, E^S_2, E^S_3]$ contains
one even and two odd numbers, with the even number having color $a$, then $S$ belongs to class $S_a$. 
If they contain one odd and two even numbers, with the odd number having color $a$,
then $S$ belongs to  class $S_a'$. In both cases, the propagator must have color $a$ to satisfy the topological
requirement, so 
\be
m_{S_a}^2=m_{S'_a}^2=m_a^2, \quad m^2_{T_0}=M_0^2,\quad m^2_{T_a}=M_a^2,\quad (a=1, 2, 3).
\labels{m22}\ee

\subsection{Complementary condition}
According to \eq{tr2}, the whole diagram must have $[E_1, E_2, E_3]$  all even or all odd, and  $E_a+E_b\ge 2$
must be fulfilled for every pair of color. These topological requirements can guarantee $m_S^2=m_{\overline S}^2$ only
when all the masses in \eq{m22} are equal.

To see that, suppose $S=T_0$, then the topological requirement merely demands $\overline S$ to be either $T_0$, or $T_a$
if $E_a(\overline S)$ is even. Thus $M_0^2=M_a^2$ for all $a$. If  $S=S_a$, then $\overline S$ could be in class $S_a$, $S_a'$, or class
$T_a$ if $E_a(\overline S)$ is odd. Hence $m_a^2=M_a^2$. Thus all the masses in \eq{m22} are equal.

\subsection{$\boldsymbol{\f_1\f_2\f_3}$ vs $\boldsymbol{\f^3}$} 
With all the masses in \eq{m22} equal, the only difference between the $\f_1\f_2\f_3$ and the $\f^3$ amplitudes
is that some Feynman diagrams allowed in $\f^3$ would not be allowed in $\f_1\f_2\f_3$. These
are the diagrams violating the topological requirements discussed in Sec.~VA.

\section{Composite Theories}
Once a set of $\m_{ij}$ satisfying the complementary condition \eq{mcomp}, the symmetry condition \eq{mcond0}, and the covariant condition
\eq{mcond1} is given, a
\M invariant scalar amplitude can be computed from \eq{chy} and 
\eq{In}. Since these three conditions  are all linear in $\m_{ij}$, 
a linear combination of two such
sets of $\m_{ij}$ also satisfies the three conditions. In this way we can construct many composite theories 
from two or more known theories by making linear combinations. In this section we shall  illustrate
how to do that with a simple example.
Clearly many other examples can  be similarly constructed.

Let $\m_{ij}=\a \m'_{ij}+\b \m''_{ij}$, where $\m'_{ij}$ comes from a Yukawa theory, with nucleon mass $m_1'$
and pion mass $m_0'$, and $\m''_{ij}$ comes from another Yukawa theory, with nucleon mass $m_1''$
and pion mass $m_0''$. $\a$ and $\b$ are arbitrary real parameters. 

Let $\n'=0, 1$ and $\n''=0, 1$  denote the nucleon number of the particles in the two
 basic theories. Then there are four kinds of particles in the composite theory: $(\n', \n'')=(0, 0), (1, 0), (0, 1), (1, 1)$.
 We can interpret $\pi=(0, 0)$ as a new (composite) pion with some mass (square) $m_0^2$, 
 $N_1=(1, 0)$ as a new nucleon with mass $m_1^2$, $N_2=(0,1)$ as a second new nucleon with mass $m_2^2$,
 and $N^*=(1,1)$ as a new di-nucleon resonance of mass $M^2$.

Fig.8(a) gives an example of a scattering diagram of the composite theory expressed in lines and vertices of the two basic theories.
It is obtained by superimposing a  diagram from the second basic theory on Fig.5(a) of the first
basic theory. As in Fig.5(a), the nucleon and pion of the first theory are depicted by thin solid and dotted lines.
The nucleon and pion of the second theory are depicted by thicker solid and dashed lines.
Fig.8(b) is the same diagram expressed in terms of the composite  particles $\pi, N_1, N_2$, and $N^*$, and 
the effective vertices  $N_1N_2N^*$ and $N^*N^*\pi$. Dotted lines are composite pions $\pi$, and thin, medium, thick solid
lines are the composite nucleons $N_1, N_2$ and the resonance $N^*$.
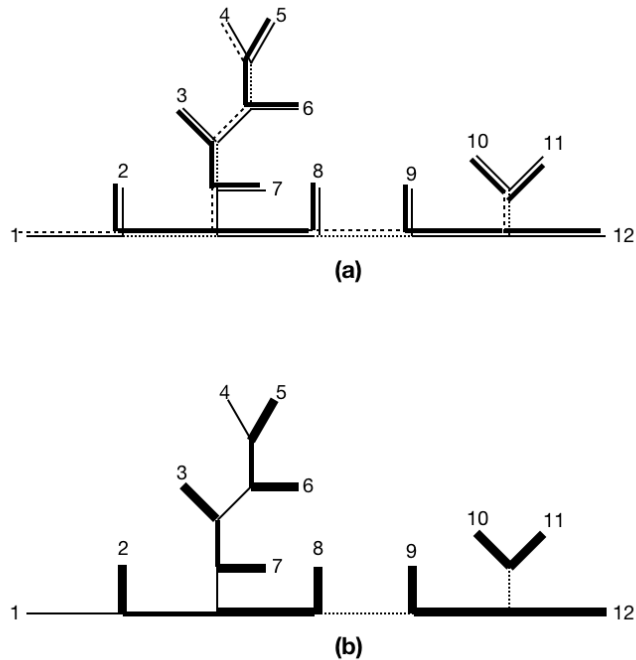
\bc\igw{10}{Fig8}\\ Fig.8\quad A Feynman diagram of a double-Yukawa composite theory. (a) is drawn
in terms of the two underlying theories, and (b) is drawn in terms of the composite particles. A dotted
line in (b) is the new pion $\pi$, a thin, medium, and thick solid line represent respectively $N_1, N_2$, and $N^*$.  \ec

The propagator masses are
\be
m_S^2&=&\a\mp_0^2+\b\mpp_0^2=m_0^2,\quad (\n', \n'')=({\rm even,\ even}),\nn\\
&=&\a\mp_0^2+\b\mpp_1^2=m_2^2,\quad (\n', \n'')=({\rm even,\ odd}),\nn\\
&=&\a\mp_1^2+\b\mpp_0^2=m_1^2,\quad (\n', \n'')=({\rm odd,\ even}),\nn\\
&=&\a\mp_1^2+\b\mpp_1^2=M^2,\quad (\n', \n'')=({\rm odd,\ odd}).\labels{mcompm}\ee
By adjusting the values of $\a, \b, \mp_0^2, \mp_1^2, \mpp_0^2, \mpp_1^2$, we can have any value
for the compound masses $m_0^2, m_1^2, m_2^2$, and $M^2$.

As a direct check, Table I shows the nine propagator masses in Fig.8(b), obtained from this formula
using the total $(\n', \n'')$ values read off  from all the corresponding external
lines in Fig.8(a). They agree with what  Fig.8(b) shows.

$$\ba{|c|c|c|c|c|c|}\hline
{\rm external\ lines}&12&45&456&3456&34567\\
(\n', \n'')&(2,1)&(2,1)&(3,2)&(4,3)&(5,4)\\
m_S^2&m_2^2&m_2^2&m_1^2&m_2^2&m_1^2\\ \hline
{\rm external\ lines}&1234567&(10)(11)&(10)(11)(12)&9(10)(11)(12)&\\
(\n', \n'')&(7,5)&(2,2)&(3,3)&(4,4)&\\
m_S^2&M^2&m_0^2&M^2&m_0^2&\\
\hline\ea$$
\bc Table I.\quad Propagator masses $m_S^2$ obtained from \eq{mcompm} and Fig.8(a) can be seen
to be the same as those shown in Fig.8(b)\ec

\section{Scalar QED}
The scattering function $f_i$ for scalar QED is identical to that of a Yukawa theory in Sec.~III, provided we interpret $m_1\equiv m$ as the
mass of the charged particle, and $m_0=0$ as the mass of photon. However,  Feynman amplitudes
in QED also contain non-trivial numerators describing photon polarizations as well as its derivative coupling to the 
charged particles. To implement that, the ${\cal I}_n$ factor in \eq{chy} must be different from \eq{In}. For an amplitude with
two  charged  legs and
and $n-2$ photon legs, we propose to use
\be {\cal I}_n&=&{N\over\s_{(12\cdots n )}},\qquad {\rm where}\nn\\
N&=&\pf'(\Psi)={(-1)^{i+j}\over\s_{ij}}\pf(\Psi^{ij}_{ij}).\labels{In2}\ee
 $\Psi$ is the $2n\x 2n$ antisymmetric matrix
\be
\Psi={\small \bm 0&0&\cdots& 0&-C_{11}&-C_{21}&\cdots& -C_{n1}\\
               0&0&\cdots& 0&-C_{12}&-C_{22}&\cdots &-C_{n2}\\
               \vdots&\vdots&&\vdots&\vdots&\vdots&&\vdots\\
               0&0&\cdots& 0&-C_{1n}&-C_{2n}&\cdots &-C_{nn}\\
               C_{11}&C_{12}&\cdots&C_{1n}&0&0&\cdots&-1/\s_{n1}\\
               C_{21}&C_{22}&\cdots&C_{2n}&0&0&\cdots&0\\
               \vdots&\vdots&&\vdots&\vdots&\vdots&&\vdots\\
               C_{n1}&C_{n2}&\cdots&C_{nn}&1/\s_{n1}&0&\cdots&0\\
               \em} ,\labels{psi}\ee
and $\Psi^{ij}_{ij}$ is $\Psi$ with the $ith$ and $j$th columns and rows removed. The matrix elements  are
\be C_{ij}&=&{\e_i\.k_i\over\s_{ij}}:={c_{ij}\over\s_{ij}},\quad (i\not=j),\nn\\
C_{ii}&=&-\sum_{j\not=i}C_{ij},\labels{cij}\ee
where $\e_i$ is the polarization vector of photon $i$. Particles 1 and $n$ are the charged scalars, the others are photons.

The motivation of this proposal is explained in Appendix C. In the rest of this section, the proposal will be explicitly 
verified for Compton scattering ($n=4$)
and for the process with a single photon emission ($n=5$), by comparing the result of \eq{chy} with the Feynman amplitude obtained
from Feynman rules.

\subsection{Compton scattering}
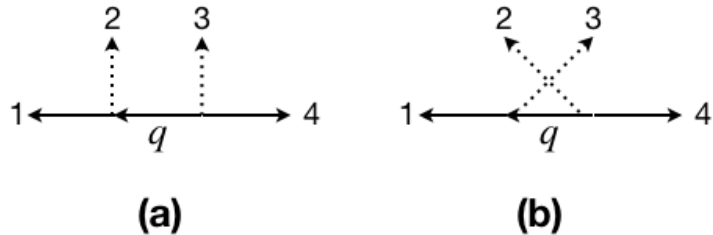
\bc\igw{10}{Fig9}\\ Fig.9\quad Compton scattering diagrams. Solid and dotted lines are charged particles and photons, respectively\ec

\subsubsection{Feynman amplitude}
Up to normalization, the Feynman amplitude of Fig.9(a) is
\be A^a_F={\e_2.(-q-k_1)\ \e_3\.(k_4-q)\over (k_3+k_4)^2-m^2}=-4{\e_2\.k_1\ \e_3\.k_4\over (k_3+k_4)^2-m^2}.\labels{f4}\ee
To be gauge invariant, we must also add in the cross diagram shown in Fig.9(b), which is the same as Fig.9(a) with $k_2$ and $k_3$ as well as $\e_2$ and $\e_3$ interchanged.

\subsubsection{CHY amplitude}
From \eq{psi} and \eq{cij} for $n=4$, one gets
\be  N&=&-{1\over\s_{13}\s_{41}}\({c_{24}c_{32}\over\s_{24}\s_{32}}+{c_{21}c_{34}\over\s_{21}\s_{34}}+{c_{23}c_{34}\over\s_{23}\s_{34}}+{c_{24}c_{34}\over\s_{24}\s_{34}}\).\labels{n4}\ee
Substituting this into \eq{In2}, and choose $\s_{r,s,t}$ to be $\s_{1,2,3}$ in \eq{chy},
the CHY amplitude can be computed  following the procedure outlined in Sec.~IIE. Instead of  poles from $1/\s_{(\a)}$, we must now look for poles in $N$ in the integration variable $\s_4$. Since line 2 is not neighboring to line 4, poles from
$\sim 1/\s_{42}$ do not count. Moreover,
as discussed in Sec.~IIID, we must also exclude poles that can lead to a would-be propagator of two pions. When applied to Fig.9(a), this means
that we should also ignore poles in $N$ proportional to $1/\s_{41}$, leaving behind only poles of the form $1/\s_{43}$ to be
considered. Thus the first
term in \eq{n4} does not contribute. For the remaining three terms, the residue of $1/\s_{43}$ is proportional to
the propagator $1/[(k_4+k_3)^2-m^2]$.
Any remaining $\s_4$ after the integration should be replaced by $\s_3$ because the integration region comes from
the vicinity of $\s_{43}=0$. Up to a possible normalization factor, the CHY amplitude from \eq{chy} is then
\be
A^a_{CHY}={\s_{(123)}\over(k_4+k_3)^2-m^2}\({1\over\s_{13}\s_{31}}\)\({c_{21}c_{34}\over\s_{21}}+{c_{23}c_{34}\over\s_{23}}+{c_{24}c_{34}\over\s_{23}}\),\nn\ee
in which the $\s_{(123)}=\s_{(rst)}$ factor comes from the integrand of \eq{chy}. Using the momentum conservation 
relation $c_{21}=-(c_{23}+c_{24})$,
we end up with
\be A^a_{CHY}={c_{21}c_{34}\over (k_4+k_3)^2-m^2},\labels{a4chy}\ee
which agrees with the Feynman amplitude $A^q_f$ in \eq{f4} up to a normalization factor. The correctness of the prescription
\eq{In2} for $n=4$ is thereby verified.

The cross diagram Fig.9(b) can be obtained either by interchanging 2 and 3 in \eq{a4chy}, or by changing ${\cal I}_n$ in \eq{In2} to $N/\s_{(1324)}$.

\subsection{Photon emission}
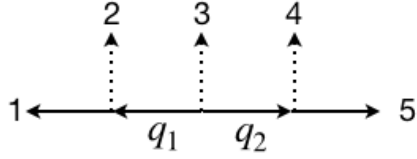
\bc\igw{6}{Fig10}\\ Fig.10\quad A tree diagram for a single photon emission in photon proton scattering. Solid and dotted lines are charged particles and photons, respectively\ec
\subsubsection{Feynman amplitude}
Up to normalization, the Feynman amplitude of Fig.10 is
\be A_F&=&{\e_2.(-q_1-k_1)\ \e_3\.(q_2-q_1)\ \e_4\.(k_5+q_2)\over[ (k_3+k_4)^2-m^2]\ [ (k_4+k_5+k_5)^2-m^2]}\nn\\
&=&-4{\e_2\.k_1\ \e_3\.(k_4+k_5-k_1-k_2)\ \e_4\.k_5\over [ (k_4+k_5)^2-m^2]\ [ (k_3+k_4+k_5)^2-m^2]}\nn\\
&=&-8{c_{21}(c_{34}+c_{35})c_{45}\over  [ (k_4+k_5)^2-m^2]\ [ (k_3+k_4+k_5)^2-m^2]}.\labels{f5}\ee
Again for gauge invariance, we must add in five other cross diagrams obtained by permuting 2, 3, 4, but we will not consider them here.

\subsubsection{CHY amplitude}
From \eq{In2} and \eq{psi} for $n=5$, one gets
{\small
\be  N&=&-{c_{2 5} c_{3 2} c_{4 1}\over \s_{1 3} \s_{2 
    5} \s_{3 2} \s_{4 1} \s_{5 1}} - {
 c_{2 1} c_{3 5} c_{4 1}\over \s_{1 3} \s_{2 1} \s_{
   3 5} \s_{4 1} \s_{5 1}} - {
 c_{2 3} c_{3 5} c_{4 1}\over \s_{1 3} \s_{2 3} \s_{
   3 5} \s_{4 1} \s_{5 1}} - {
 c_{2 4} c_{3 5} c_{4 1}\over \s_{1 3} \s_{2 4} \s_{
   3 5} \s_{4 1} \s_{5 1}} - {
 c_{2 5} c_{3 5} c_{4 1}\over \s_{1 3} \s_{2 5} \s_{
   3 5} \s_{4 1} \s_{5 1}}\nn\\
   && - {
 c_{2 5} c_{3 2} c_{4 2}\over \s_{1 3} \s_{2 5} \s_{
   3 2} \s_{4 2} \s_{5 1}} - {
 c_{2 5} c_{3 4} c_{4 2}\over \s_{1 3} \s_{2 5} \s_{
   3 4} \s_{4 2} \s_{5 1}} - {
 c_{2 1} c_{3 5} c_{4 2}\over \s_{1 3} \s_{2 1} \s_{
   3 5} \s_{4 2} \s_{5 1}} - {
 c_{2 3} c_{3 5} c_{4 2}\over \s_{1 3} \s_{2 3} \s_{
   3 5} \s_{4 2} \s_{5 1}} - {
 c_{2 5} c_{3 5} c_{4 2}\over \s_{1 3} \s_{2 5} \s_{
   3 5} \s_{4 2} \s_{5 1}} \nn\\
   &&- {
 c_{2 5} c_{3 2} c_{4 3}\over \s_{1 3} \s_{2 5} \s_{
   3 2} \s_{4 3} \s_{5 1}} - {
 c_{2 1} c_{3 5} c_{4 3}\over \s_{1 3} \s_{2 1} \s_{
   3 5} \s_{4 3} \s_{5 1}} - {
 c_{2 3} c_{3 5} c_{4 3}\over \s_{1 3} \s_{2 3} \s_{
   3 5} \s_{4 3} \s_{5 1}} - {
 c_{2 4} c_{3 5} c_{4 3}\over \s_{1 3} \s_{2 4} \s_{
   3 5} \s_{4 3} \s_{5 1}} - {
 c_{2 5} c_{3 5} c_{4 3}\over \s_{1 3} \s_{2 5} \s_{
   3 5} \s_{4 3} \s_{5 1}} \nn\\
   &&- {
 c_{2 4} c_{3 2} c_{4 5}\over \s_{1 3} \s_{2 4} \s_{
   3 2} \s_{4 5} \s_{5 1}} - {
 c_{2 5} c_{3 2} c_{4 5}\over \s_{1 3} \s_{2 5} \s_{
   3 2} \s_{4 5} \s_{5 1}} - {
 c_{2 1} c_{3 4} c_{4 5}\over \s_{1 3} \s_{2 1} \s_{
   3 4} \s_{4 5} \s_{5 1}} - {
 c_{2 3} c_{3 4} c_{4 5}\over \s_{1 3} \s_{2 3} \s_{
   3 4} \s_{4 5} \s_{5 1}} - {
 c_{2 4} c_{3 4} c_{4 5}\over \s_{1 3} \s_{2 4} \s_{
   3 4} \s_{4 5} \s_{5 1}} \nn\\
   &&- {
 c_{2 5} c_{3 4} c_{4 5}\over \s_{1 3} \s_{2 5} \s_{
   3 4} \s_{4 5} \s_{5 1}} - {
 c_{2 1} c_{3 5} c_{4 5}\over \s_{1 3} \s_{2 1} \s_{
   3 5} \s_{4 5} \s_{5 1}} - {
 c_{2 3} c_{3 5} c_{4 5}\over \s_{1 3} \s_{2 3} \s_{
   3 5} \s_{4 5} \s_{5 1}} - {
 c_{2 4} c_{3 5} c_{4 5}\over \s_{1 3} \s_{2 4} \s_{
   3 5} \s_{4 5} \s_{5 1}} - {
 c_{2 5} c_{3 5} c_{4 5}\over \s_{1 3} \s_{2 5} \s_{
   3 5} \s_{4 5} \s_{5 1}}.\nn\ee}
As before, choose $\s_{r,s,t}=\s_{1,2,3}$ in \eq{chy}, then search for poles in $N$ in the integration variables $\s_4$ and $\s_5$. 
Poles with non-neighboring lines do not count, and poles leading up to pure pions propagators should also be ignored. Therefore
the relevant poles in $N$ only come from terms containing a $1/(\s_{35}\s_{45})$ factor, 
or a 
$1/(\s_{34}\s_{45})$ factor, both giving rise to a residue proportional to $1/[(k_4+k_5)^2-m^2][(k_3+k_4+k_5)^2-m^2]$. 
All left over $\s_4$ and $\s_5$ factors after the integration should be set equal to $\s_3$.

With that in mind, of the 25 terms in $N$, only  terms  18 to 25 contribute. Up to a possible normalization factor, the CHY amplitude
for $n=5$ is then
\be
A_{CHY}&=&{\s_{(123)}\ B\ c_{45}\over [(k_4+k_5)^2-m^2][(k_3+k_4+k_5)^2-m^2]},\nn\\ \nn
\\
B&=&-{1\over\s_{13}\s_{31}}\(   {
 c_{2 1} c_{3 4} \over  \s_{2 1 } }+ {
 c_{2 3} c_{3 4} \over  \s_{2 3}  } + {
 c_{2 4} c_{3 4} \over  \s_{2 3}   }+ {
 c_{2 5} c_{3 4} \over  \s_{2 3}   }\) \nn\\
   && -{1\over\s_{13}\s_{31}}\( {
 c_{2 1} c_{3 5} \over  \s_{2 1}   } + {
 c_{2 3} c_{3 5} \over  \s_{2 3}   } + {
 c_{2 4} c_{3 5} \over  \s_{2 3}   } + {
 c_{2 5} c_{3 5} \over  \s_{2 3}   }\)\nn\\
 &=&-c_{21}(c_{34}+c_{35}){1\over\s_{13}\s_{31}}\({1\over\s_{21}}-{1\over\s_{23}}\)={c_{21}(c_{34}+c_{35})\over\s_{13}\s_{21}\s_{23}},\nn\ee
in which  momentum conservation  $c_{21}=-(c_{23}+c_{24}+c_{25})$ has been used. Substituting $B$ into $A_{CHY}$,
one gets
\be
A_{CHY}={c_{21}(c_{34}+c_{35})c_{45}\over [(k_4+k_5)^2-m^2][(k_3+k_4+k_5)^2-m^2]},\labels{a5chy}\ee
which up to a normalization factor is the same as the Feynman amplitude $A_F$ in \eq{f5}, thereby verifying the correctness of \eq{In2} for $n=5$.

\section{Disk and Sphere Integrals}
A disk integral
\be Z_P(q_1q_2\cdots q_n)&=&(\a')^{n-3}\int_{D(P)}{dz_1\ dz_2\ \cdots\ dz_n\over{\rm vol}(SL(2,R))}\  {\prod_{i<j}|z_{ij}|^{\a'k_i\cdot k_j}\over z_{q_1q_2}z_{q_2q_3}\cdots z_{q_{n-1}q_n}z_{q_nq_1}}\labels{Z}\ee
was introduced in \cite{BDV09, sS09} to connect a Yang-Mills field theory amplitude with an open string amplitude. Following \cite{BSS13},
the integration variables $\s_i$ are now denoted as $z_i$.  $\a'$ is the Regge slope, and $k_i$ are the light-like external momenta. The integration domain is the real line with the integration variables ordered
according to a permutation $P=\{p_1p_2\cdots p_n\}\in S_n$,
\be D(P)&=&\{(z_1z_2\cdots z_n)\in R^n|-\infty<z_{p_1}<z_{p_2}<\cdots<z_{p_n}<\infty\}.\nn\ee
The $Z$ function depends on the permutation $P$,  and also  on another permutation $Q=\{q_1q_2\cdots q_n\}\in S_n$. 
The integrand is invariant under a SL(2,R)
transformation, which is the \M transformation with real coefficients. 

The $Z$ function possesses the following properties \cite{CMS16}:
\be
 Z_P(q_1q_2\cdots q_n)&=&Z_P(q_2q_3\cdots q_1),\labels{Z1}\\
 Z_P(q_1q_2\cdots q_n)&=&(-)^nZ_P(q_nq_{n-1}\cdots q_1),\labels{Z2}\\
 Z_P(1, A, n, B)&=& (-)^{|B|}\sum_{A\sh\tilde B}  Z_P(1,\s, n),\quad \forall A, B,\labels{Z3}\\
 0&=&\sum_{k=2}^{n-1}k_{q_1}\.(k_{q_2}\+k_{q_3}\+\cdots \+k_{q_j})Z_P(q_2q_3\cdots q_jq_1q_{j+1}\cdots q_n) ,\labels{Z4}\\
 Z_{p_1 p_2 \cdots p_n} (Q) &=& Z_{p_2 p_3 \cdots p_n p_1} (Q) = (-1)^n Z_{p_n \cdots p_2 p_1} (Q),\labels{Z5}\\
 0&=&\sum_{j=1}^{n-1}\exp\[i\pi\a' k_{p_1}\.(k_{p_2}\+k_{p_3}\+\cdots \+k_{p_j})\]Z_{p_2p_3\cdots p_jp_1p_{j+1}\cdots p_n},  \labels{Z6}\ee
where $\tilde B$ is the transpose of the set $B$, and $A\sh \tilde B$ is the shuffle product between set $A$ and set $\tilde B$.
In the zero Regge slope limit, $Z_P(Q)$ is essentially the scalar amplitude \eq{chy}, with $P=\{12\cdots n\}$ and $Q=\a$.

The $Z$ function can be generalized to massive and off-shell situations by replacing $k_i\.k_j$ in \eq{Z} with $a_{ij}$ of the previous sections.
Since $a_{ii}=0$ and $\sum_{j\not=i}a_{ij}=0$, the Koba-Nielsen factor $|z_{ij}|^{\a'a_{ij}}$ remains invariant under a \M transformation,
thereby leaving the integral in \eq{Z} well defined. This modified $Z$ function still possesses all the same properties, \eq{Z1} to \eq{Z6}, except
that the momentum dot products in \eq{Z4} and \eq{Z6} should be replaced by the sum of $a_{ij}$ factors:
\be k_{q_1}\.(k_{q_2}\+k_{q_3}\+\cdots \+k_{q_j})&\to& \sum_{j=2}^k a_{q_1q_j},\nn\\
k_{p_1}\.(k_{p_2}\+k_{p_3}\+\cdots \+k_{p_j})&\to& \sum_{j=2}^k a_{p_1p_j}.\labels{Z7}\ee
In the zero Regge slope limit, it again reduces to the field theory amplitude defined by the same $a_{ij}$.

With these similarities, it would be interesting to find out whether  modified string amplitudes and/or new effective field theories
can be sensibly constructed with the help of
the modified $Z$ function.

A similar remark applies to the modified sphere function, defined by 
\be
J_P(q_1q_2\cdots q_n)&=&(\a')^{n-3}\int_{C^n}{d^2z_1\ d^2z_2\ \cdots\  d^2z_n\over{\rm vol}(SL(2,C))}\cdot\nn\\
&&  {\prod_{i<j}|z_{ij}|^{2\a'a_{ij}}\over (z_{p_1p_2}z_{p_2p_3}\cdots z_{p_{n-1}p_n}z_{p_np_1})(z_{q_1q_2}z_{q_2q_3}\cdots z_{q_{n-1}q_n}z_{q_nq_1})}.\labels{J}\ee
When the corresponding field theory is massless and on-shell, $a_{ij}=k_i\.k_j$, the $J$ function returns to the original one in \cite{SS18}.

\section{Summary}
By modifying the scattering equations used in the CHY  tree amplitudes, we extended the formalism to many
off-shell scalar amplitudes carrying non-zero masses. The modification consists of changing $k_i\. k_j$ to an arbitrary
$a_{ij}=k_i\. k_j+\r_{ij}+\m_{ij}$, with  $\sum_{j\not=i}a_{ij}=0$. The term $\r_{ij}$ provides an off-shell extension,
and the term $\m_{ij}$ specifies how non-zero masses enter  cubic interactions. The same modification can be applied
to the Koba-Nielsen factor of disk and sphere functions, which may be useful in generalizing string related amplitudes. 
Future work includes how amplitudes involving
higher spins can be similarly constructed, and how  off-shell extension can be used to study amplitudes with  any
number of loops.

\appendix
\section{Masses and off-shell momenta coming from extra dimensions}
Consider an $n$-particle scattering amplitude with external momenta $\hat k_i$ in a higher dimensional space, and
suppose $\hat k_i^2=0$ for all $i$.
Decompose $\hat k_i=(k_i, k_i')$ into a four-dimensional component $k_i$
and an extra-dimensional component $k_i'$. Since
$k_i^2=-k_i^{'2}$ may take on any value,
 a non-zero mass $m$ and/or an off-shell four-momentum can be produced 
 in real space-time. We show in this Appendix that masses  produced this way 
 do not {\it always} yield the right propagators in a massive theory, though it might do so
 if only certain particles  in the amplitude are massive. Similar arguments apply also to off-shell momenta.
   
Consider a theory whose particles carry a mass $m$. 
 Label the external lines cyclically in the natural order $i=1, 2, \cdots, n$. For any $p$ and any $q\le n-2$,
there is always an inverse
propagator  $(\sum_{i=p}^{p+q-1}k_i)^2-m^2$ present in some tree diagram. If masses 
 come strictly from the extra dimensional momenta $\hat k_i$, then such an inverse propagator 
must be $(\sum_{i=p}^{p+q-1}\hat k_i)^2 = 
(\sum_{i=p}^{p+q-1} k_i)^2+(\sum_{i=p}^{p+q-1} k'_i)^2$. This requires 
\be \(\sum_{i=p}^{p+q-1} k'_i\)^2=-m^2\labels{A1} \ee
for any $p$ and any $q\le n-2$. This turns out to be impossible for a fixed extra dimension and a large enough
$n$.

Using \eq{A1} and $ k_p^{'2}=-m^2$, we get
\be 2k_p'\.\(\sum_{i=p+1}^{p+q-1}k'_i\)=m^2\labels{A2}\ee
for any $2\le q\le n-2$. Hence $k'_p\.k'_i=0$ for all $i\ge p+2$ as long as $i-p+1\le n-2$. This cannot happen
if $n$ is large enough, because $k'_p\.k'_i=0$ can occur only for $(d-1)$ $i$'s in a $d-$dimensional extra dimension.

This concludes the proof that we cannot obtain all propagators of all Feynman diagrams
correctly if masses for a massive field theory 
come solely from the extra dimensions.

\section{Covariance and sum rules}
If $a_{ij}$ satisfies the constraints $a_{ii}=0$ and $\sum_ja_{ij}=0$, then
$\hat f_i(\s)$ will transform covariantly as in \eq{cov}, and the sum rules \eq{sr} will be satisfied.

To prove these assertions, note that under a \M transformation given by \eq{M}, $1/(\s-\s_i)\to\l\l_i/(\s-\s_i)$ where $\l=(\g\s+\d)$ and
$\l_i=(\g\s_i+\d)$. Hence
\be
\hat f_i(\s)&=&\sum_{j=1,j\not=i}{a_{ij}\over\s-\s_j}\to \l\sum_{j\not=i}{a_{ij}\l_j\over\s-\s_j}
=\l\sum_{j\not=i}{a_{ij}\over\s-\s_j}\[\l-\g(\s-\s_j)\]=\l^2\hat f_i(\s)\nn\ee
because $\sum_ja_{ij}=0$. 

For the sum rules, recall that $f_i=\hat f_i(\s_i)=\sum_{j\not=i}a_{ij}/(\s_i-\s_j)$, hence
\be
\sum_if_i&=&\sum_{i\not=j}{a_{ij}\over\s_i-\s_j}=0,\nn\\
\sum_if_i\s_i&=&\sum_{i\not=j}{a_{ij}\s_i\over\s_i-\s_j}=\h\sum_{i\not=j}{a_{ij}(\s_i-\s_j)\over\s_i-\s_j}=0,\nn\\
\sum_if_i\s_i^2&=&\sum_{i\not=j}{a_{ij}\s_i^2\over\s_i-\s_j}=\sum_{i\not=j}{a_{ij}\s_i(\s_i-\s_j)\over\s_i-\s_j}
+\sum_{i\not=j}{a_{ij}\s_i\s_j\over\s_i-\s_j}=0.
\nn\ee
The last of the first equation is 0 because the summand is antisymmetric in $i,j$. The last term of the second equation is
zero because $\sum_ja_{ij}=0$. The first term of the last expression of the last equation is zero because $\sum_ja_{ij}=0$, 
and the second term of that expression
is zero because the summand is antisymmetric in $i$ and $j$.

\section{Numerator factor for QED}
The  numerator factor $N$ in \eq{In2}  comes from a comparison between the Feynman
rules for QED and those for gluon scattering. 
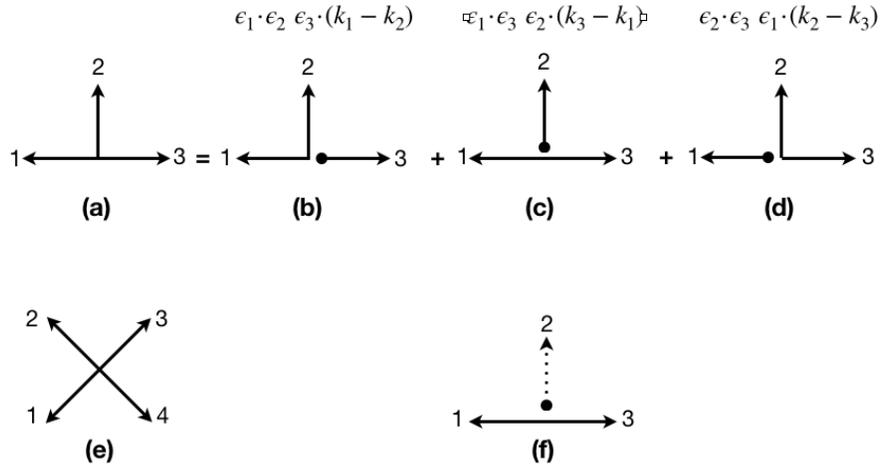
\bc\igw{12}{Fig11}\\ Fig.11.\quad Vertex factors for QCD shown in (a) and (e), and for scalar QED shown in (f)\ec

For gluon scatterings, there is the four-gluon vertex in Fig.11(e), and  the three-gluon vertex Fig.11(a), 
containing three terms shown separately in
diagrams (b), (c), and (e). The vertex factor for Fig.11(c) is identical to the vertex factor in scalar QED, shown in Fig.11(f), except for $\e_1\.\e_3$.

A gluon amplitude is a function of $\e_i\.\e_j,\ \e_i\. k_j$, and $k_i\.k_j$. In the CHY theory \cite{CHY13b, CHY13c}, the numerator 
of the amplitude comes from the reduced Pfaffian of a $2n\x2n$ antisymmetric matrix of the form
\be \Psi=\bm A&-C^T\\ C&B\\ \em, \labels{qcd}\ee
where for $i\not=j$, $A_{ij}=k_i\.k_j/\s_{ij},\ C_{ij}=\e_i\.k_j/\s_{ij}$, and $B_{ij}=\e_i\.\e_j/\s_{ij}$.
A scalar QED amplitude with two charged scalars occupying lines 1 and $n$, such as Fig.9 or Fig.10, contains $\e_i\.k_j$ factors but no $\e_i\.\e_j$ nor $k_i\.k_j$
factors in the numerator. Since Fig.11(c) is essentially the same as Fig.11(f), we may obtain the numerator factor $N$ in such QED
diagrams from \eq{qcd}
by putting $A=0$, and $B=0$ except for the $\e_1\.\e_n/\s_{1n}$ and the $\e_n\.\e_1/\s_{n1}$ elements.
This is then the $N$ factor in \eq{In2}.

There is however a caveat in using $N$. A number of gluons can merge to form a gluon propagator, but photons in a tree diagram cannot
merge to form photon propagators. Hence those poles in $N$ that lead to the phantom photon propagators must be excluded by hand.
Other than this, $N$ has another imperfection. The reduced Pfaffian
for the gluon matrix in \eq{qcd} is gauge invariant, but this is not so for the photon matrix in \eq{psi}.

Moreover, this argument to produce $N$ no longer works when two pairs of charged legs are present, for two different reasons. 
First, there would now be two  $\e_i\.\e_j$ factors in the $B$ block for the two charged pairs.
If the factor $B$ in \eq{qcd} contains two such pairs of $\e_i\.\e_j$, then their product  would appear somewhere
in the reduced Pfaffian. In the case of gluon scattering, this product could come from the four-gluon vertices Fig.11(e),
or terms such as Fig.11(b) or Fig.11(d).  In the case of scalar QED, all these diagrams are absent, so this term would be extraneous and wrong. 
The second reason why that does not
work is because   $k_i\.k_j$ factors would appear when the two pairs exchange a photon, 
so the $A$ block in \eq{qcd} can no longer be dropped.
However,  \M invariance demands $A$ to be 
$a_{ij}/\s_{ij}$ when masses are present \cite{Lam19}, not just $k_i\.k_j/\s_{ij}$. Then mass factors $\m_{ij}/\s_{ij}\sim m/\s_{ij}$ 
would also appear in $A$ and in the numerator of the scattering amplitude. That is wrong and should not happen.
For both of these reasons, the $N$ factors for scalar QED can no longer be
obtained so simply from the gluon $N$ factor when more than one pair of charged particles are present.

\end{document}